\documentclass[fleqn,usenatbib]{mnras}

\usepackage[T1]{fontenc}
\usepackage{ae,aecompl}
\usepackage{txfonts}
\usepackage{mathptmx}
\usepackage{graphicx}
\usepackage{amssymb}
\usepackage{aas_macros}
\usepackage{color}
\usepackage{hyperref}

\title[Verification of a photometric PSB selection]{The identification of post-starburst galaxies at $z\sim1$
using multiwavelength photometry: a spectroscopic verification
\vspace{-0.0cm}}

\author[D.~T.~Maltby et al.]
{David~T.~Maltby$^{1}$\thanks{E-mail: dtmaltby@gmail.com},
 Omar~Almaini$^{1}$, Vivienne~Wild$^{2,3}$, Nina~A.~Hatch$^{1}$,
\newauthor William~G.~Hartley$^{4}$, Chris~Simpson$^{}$,
Ross~J.~McLure$^{3}$, James~Dunlop$^{3}$,
\newauthor Kate~Rowlands$^{2}$ and Michele~Cirasuolo$^{5}$\\
$^{1}$School of Physics and Astronomy, University of Nottingham, University Park, Nottingham, NG7 2RD, UK\\
$^{2}$School of Physics and Astronomy, University of St Andrews, North Haugh, St Andrews, KY16 9SS, UK (SUPA)\\
$^{3}$Institute for Astronomy, University of Edinburgh, Royal Observatory, Blackford Hill, Edinburgh, EH9 3HJ, UK (SUPA)\\
$^{4}$ETH Z{\"u}rich, Institut f{\"u}r Astronomie, Wolfgang-Pauli-Str. 27, CH-8093 Z{\"u}rich, Switzerland\\
$^{5}$UK Astronomy Technology Ctr., Royal Observatory, Blackford Hill, Edinburgh, EH9 3HJ, UK}

\date{Accepted yyyy mm dd. Received yyyy mm dd; in original form yyyy mm dd}

\pubyear{2016}

\begin{document}

\label{firstpage}

\pagerange{\pageref{firstpage}--\pageref{lastpage}}

\maketitle


\begin{abstract}
Despite decades of study, we still do not fully understand why some massive galaxies abruptly switch off
their star formation in the early Universe, and what causes their rapid transition to the red sequence.
Post-starburst galaxies provide a rare opportunity to study this transition phase, but few have currently
been spectroscopically identified at high redshift ($z>1$).  In this paper we present the spectroscopic
verification of a new photometric technique to identify post-starbursts in high-redshift surveys.  The
method classifies the broad-band optical--near-infrared spectral energy distributions (SEDs) of
galaxies using three spectral shape parameters (super-colours), derived from a principal component analysis
of model SEDs.  When applied to the multiwavelength photometric data in the UKIDSS Ultra Deep Survey
(UDS), this technique identified over $900$ candidate post-starbursts at redshifts $0.5<z<2.0$.  In this
study we present deep optical spectroscopy for a subset of these galaxies, in order to confirm their
post-starburst nature.  Where a spectroscopic assessment was possible, we find the majority
($19$/$24$ galaxies; $\sim 80$ per cent) exhibit the strong Balmer absorption (H\,$\delta$
equivalent width $W_{\lambda}>5$\,\AA) and Balmer break, characteristic of post-starburst galaxies.
We conclude that photometric methods can be used to select large samples of recently-quenched
galaxies in the distant Universe.
\end{abstract}

\begin{keywords}
methods: statistical ---
galaxies: fundamental parameters ---
galaxies: high-redshift ---
galaxies: photometry ---
galaxies: statistics ---
galaxies: stellar content.
\vspace{-0.4cm}
\end{keywords}

\section[]{Introduction}

\label{Introduction}

Deep surveys have transformed our view of the distant Universe, allowing us to observe the process of galaxy
assembly over the last $13$ billion years of cosmic history.  However, despite much progress, many crucial
aspects of galaxy formation and evolution remain poorly understood.  In particular, we still do not
understand the mechanisms responsible for quenching star formation in massive galaxies, as required to
produce both the `red nuggets' observed at high redshift \cite[e.g.][]{Trujillo_etal:2006} and the quiescent
galaxies we observe today \cite[e.g.][]{Baldry_etal:2004}.

To terminate star formation in galaxies, several mechanisms have been proposed.  These include, gas stripping
\citep[e.g.][]{Gunn&Gott:1972}, morphological quenching \citep{Martig_etal:2009}, AGN or starburst-driven
superwinds \citep{Hopkins_etal:2005, Diamond-Stanic_etal:2012}, shock heating of infalling cold gas by the
hot halo \citep{Dekel&Birnboim:2006}, and an exhaustion of the gas supply \citep[e.g.][]{Larson_etal:1980}.
Furthermore, to keep star formation suppressed, radio-mode AGN feedback may also be required
\break\citep{Best_etal:2005}.

To disentangle the relative roles of these processes at high redshift, we require a large sample of galaxies
that are caught in the act of transformation.  The rare class of `post-starburst' galaxies is potentially
ideal for such studies.  Otherwise known as `E+A' or `k+a' galaxies, post-starbursts represent transition
systems in which a major burst of star formation was quenched within the last few hundred Myr, leaving a
characteristic A-star imprint of strong Balmer lines on an otherwise passive-looking galaxy spectrum
\citep{Dressler&Gunn:1983,Wild_etal:2009}.  However, due to their very red spectral energy distributions
(SEDs) and intrinsic short-lived nature, only a handful have been spectroscopically identified at $z>1$;
e.g.\ $3$~galaxies in zCOSMOS \citep{Vergani_etal:2010} and $5$ galaxies in the spectroscopic component of
the Ultra Deep Survey \break\citep[UDSz;][]{Wild_etal:2014}.

In order to increase the number of post-starburst (PSB) galaxies identified at high redshift, two photometric
methods have recently been developed.  \cite{Whitaker_etal:2012} use medium-band infrared imaging to identify
`young red-sequence' galaxies via rest-frame {\em UVJ} colour-colour diagrams.  \cite{Wild_etal:2014},
however, use an alternative approach and classify galaxies based on a principal component analysis (PCA) of
their SEDs.  When this latter method is applied to the multiwavelength photometry of the Ultra Deep Survey
(UDS; Almaini et al., in preparation), it transpires that just three shape parameters (i.e.~eigenspectra)
provide a compact representation of a wide variety of SED shapes.  This enables the identification of several
galaxy populations, including PSBs which have formed $>10$ per cent of their mass in a recently-quenched
\break starburst event.

Using photometry for over $130\,000$ $K$-band selected UDS galaxies, this PCA method has identified over
$900$ candidate PSBs at $0.5 < z < 2.0$ (Wild et al., in preparation).  Such a large sample of high-redshift
post-starbursts will enable a wide range of scientific studies and provide crucial insight into the quenching
of star formation in galaxies at high redshift.  At present, however, the spectroscopic confirmation rests on
a limited number of low-resolution spectra from UDSz \citep[$5$ PSBs;][]{Wild_etal:2014}.  In this study, we
provide a more robust assessment using medium-resolution deep-optical spectroscopy for a much larger sample
of PSB candidates.  This provides the crucial validation required to unlock a wealth of science on
high-redshift post-starbursts selected from the UDS and other photometric surveys.

The structure of this paper is as follows.  In Section~\ref{Description of the data}, we give a brief
description of the UDS data and VIMOS spectroscopy upon which this work is based, outlining the photometric
identification of PSB candidates and our selection of spectroscopic targets in Section~\ref{Target
selection}.  In Section~\ref{Results}, we present the reduced VIMOS spectra for PSB candidates and classify
the galaxies using standard spectroscopic criteria.  Finally, we draw our conclusions in
Section~\ref{Conclusions}.

\vspace{-0.3cm}

\section[]{Description of the data}

\label{Description of the data}

\subsection[]{The Ultra Deep Survey (UDS)}

\label{The UDS}

The UDS is the deepest component of the UKIRT (United Kingdom Infra-Red Telescope) Infrared Deep Sky Survey
\cite[UKIDSS;][]{Lawrence_etal:2007} and the deepest degree-scale near-infrared survey to date\footnote{
http://www.nottingham.ac.uk/astronomy/UDS/}.  The survey comprises extremely deep UKIRT {\em JHK} photometry,
covering an area of $0.77\rm\deg^{2}$.  The work presented here is based on the eighth UDS data release (DR8)
where the limiting depths are $J = 24.9$; $H = 24.4$ and  $K = 24.6$ (AB; $5\sigma$ in $2\rm\,arcsec$
apertures).  In addition, the UDS is complemented by extensive multiwavelength observations including
deep-optical {\em BVRi$'$z$'$} photometry from the Subaru--{\em XMM-Newton} Deep Survey \citep[SXDS;\break][]
{Furusawa_etal:2008} and mid-infrared coverage ($3.6$ and $4.5\rm\,\mu{m}$) from the {\em Spitzer} UDS Legacy
Program (SpUDS; PI: Dunlop).  Furthermore, deep $u'$-band photometry is also available from the
Canada--France--Hawaii Telescope (CFHT) Megacam.  The complete extent of the UDS field with full
multiwavelength coverage (optical--mid-infrared) is $\sim0.62\rm\,deg^{2}$.  For a complete description of
these data, including a description of the catalogue construction, see \cite{Hartley_etal:2013} and
\cite{Simpson_etal:2012}.  In this work, where appropriate, we use the photometric redshifts described in
\cite{Simpson_etal:2013}.

\vspace{-0.3cm}

\subsection[]{The super-colour method and target selection}

\label{Target selection}

To identify high-redshift PSB candidates in the UDS, we use the PCA technique of \cite{Wild_etal:2014}.  We
provide a brief description of the method below.

\cite{Wild_etal:2014} begin by building a large library of $\sim44\,000$ `stochastic burst' model SEDs,
generated from \cite{Bruzual&Charlot:2003} stellar population synthesis models with stochastic star-formation
histories.  These model SEDs are convolved with the photometric filter set of the UDS and progressively
shifted to cover the redshift range of interest (e.g.\ $0.5 < z < 2.0$).  A PCA analysis is then applied to all
model SEDs in order to establish a mean SED ($m_{\lambda}$) and a series of $p$ eigenvectors (eigenspectra;
$e_{i\lambda}$) from which any normalised SED ($f_{\lambda}/{n}$) can be approximately reconstructed:
\begin{equation}
\frac{f_{\lambda}}{n} = m_{\lambda} + \sum_{i=1}^{p} a_i e_{i\lambda}.
\end{equation}

The principal component amplitudes ($a_{i}$), indicate the `amount' of each eigenspectrum contained within a
galaxy's SED and can be used to uniquely define its shape.  Referred to as {\em super-colours} (SCs), these
amplitudes combine data from multiple filters using an optimally defined weighting scheme (the eigenspectra)
and allow for all key information available from multiwavelength photometry to be retained.  In fact, it
transpires that only the first three SCs ($a_{1}={\rm SC1}$; $a_{2}={\rm SC2}$; $a_{3}={\rm SC3}$) are
required in order to provide a compact representation of a wide variety of SED shapes, accounting for $>99.9$
per cent of the variance in the models of \cite{Wild_etal:2014}.  For these SCs, several useful correlations
exist with the star-formation histories of the model SEDs.  For example, SC1 correlates with both mean
stellar age and dust content; SC2  and SC3 correlate with metallicity; and SC2 correlates with the fraction
of stellar mass formed in bursts in the last Gyr.  These key correlations enable the separation of a tight
red-sequence from star-forming galaxies and also help identify three unusual populations: very dusty
star-forming galaxies, metal poor quiescent dwarf galaxies, and PSBs that have formed $>10$ per cent of their
mass in a recently-quenched starburst (see \citealt{Wild_etal:2014}).

For real galaxies, a PCA analysis is used to project their SEDs into SC space, with a comparison to the SCs
of the model SEDs used in order to determine their probable nature (e.g.\ red-sequence, star-forming, PSB).
The benefit of this approach is that no model SED or K-correction is assumed in determining the SCs of the
actual galaxies themselves.  In the UDS, this PCA analysis was performed using
$8$ UDS filters ({\em VRi$'$z$'$JHK}, $3.6\rm\,\mu{m}$), and on all galaxies with $K_{\rm AB} < 24$ and
$0.5 < z < 2.0$ ($48\,713$ galaxies; Wild et al., in preparation).  From this analysis, a parent sample of
$4249$ red-sequence galaxies, $39\,970$ star-forming galaxies and $921$ PSB candidates have been identified.

\begin{figure}
\includegraphics[height=0.45\textwidth]{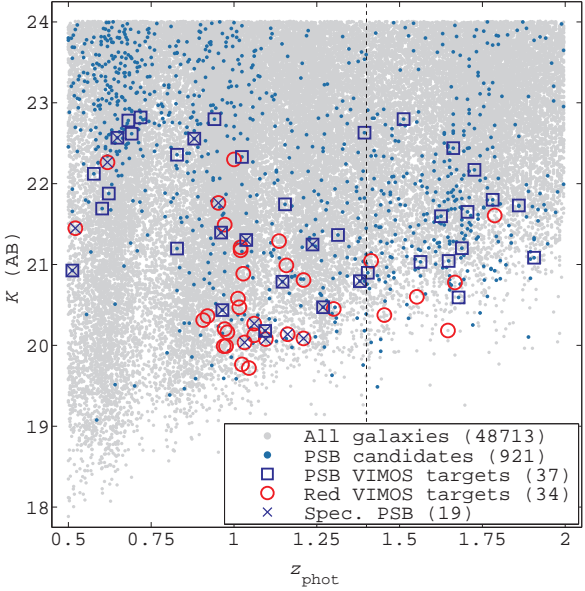}
\centering
\vspace{-0.1cm}
\caption{\label{Targets} $K$-band magnitude versus photometric redshift for UDS galaxies ($0.5 < z < 2.0$;
$K_{\rm AB} < 24$), showing the super-colour selected post-starburst (PSB) candidates (blue points).
Spectroscopic targets from our new medium-resolution spectroscopy (see
Section~\ref{VIMOS observations and data reduction}) are shown: PSB candidates (blue squares) and
red-sequence galaxies (red circles).  Respective sample sizes are shown in the legend.  Also indicated are
the upper redshift limit for our spectroscopic PSB classification ($z = 1.4$; black dashed line)
and the PSBs identified from the spectroscopy of these targets (blue crosses; see
Section~\ref{Results} and Table~\ref{UDS PSB galaxies}).  Note that spectroscopic candidates were selected
from an early version of the SC catalogue to $K_{\rm AB} < 23$.}
\end{figure}

To verify the identification of PSB galaxies, we select spectroscopic targets from this parent sample
covering a wide range in redshift and magnitude (see Fig.~\ref{Targets}).  Note, however, that selection was
based on preliminary catalogues to a limit of $K_{\rm AB} < 23$, and that targets were chosen with $i<24.5$
to ensure adequate signal-to-noise (S/N) in the optical spectra.  Red-sequence galaxies were also selected,
with priority given to those near the SC classification boundary with post-starbursts, to allow for a robust
calibration of the boundary's position.  A total of $88$ targets were selected, including $37$ PSB candidates
and a comparison sample of $34$ red-sequence galaxies.  Spectra for an additional $11$ star-forming,
$2$~dusty star-forming and $4$ low-metallicity galaxies were also taken, but these are not considered in this
paper.

\vspace{-0.3cm}

\subsection[]{VIMOS spectroscopy and data reduction}

\label{VIMOS observations and data reduction}

Spectroscopic observations were taken with the VIMOS spectrograph at the European Southern Observatory's Very
Large Telescope (ESO programme 094.A-0410).  Observations were performed in multi-object spectroscopy (MOS)
mode.  Spectroscopic data were taken with a single VIMOS mask and with total on-source exposures of
$\sim4\rm\,h$, which were divided over $7$ observing blocks (OBs) and four nights (data taken in Nov/Dec
2014).  Each individual OB had an exposure time of $\sim35\rm\,min$ ($2088\rm\,s$) and was observed in a
pattern of three $1.2$ arcsec dithered exposures.  All observations were performed in dark time, under clear
conditions, with a seeing of $<1\rm\,arcsec$ and with an airmass $<1.2$.  Standard calibration data were also
taken as part of the observing programme; including detector biases, screen flat fields, arc-lamp exposures
and spectro-photometric standards.

We use VIMOS with the medium-resolution grism (MR) and GG475 order sorting filter; a configuration that
provides a wide wavelength coverage ($4800$--$10\,000$\,\AA) and a dispersion of $2.6$\,\AA\,$\rm pixel^{-1}$.
The field of view for VIMOS is separated into four quadrants, each of which is $7\times8\rm\,arcmin^2$.  For
our MOS observations, we use slit masks with $20$--$25$ slits per quadrant, yielding spectra for a total of
$88$ galaxies in the UDS field (see Section~\ref{Target selection}).  The masks use slits of $1\rm\,arcsec$
width, resulting in a typical spectral resolution of $R = 580$.

Raw spectra were reduced using the VIMOS MOS pipeline (version 2.9.16) and within the ESO {\sc reflex}
environment \citep{Freudling_etal:2013}.  Standard data reduction procedures were followed, leading to flux-
and wavelength-calibrated spectra with a mean rms in the residuals of the wavelength calibration $<0.7$\,\AA.
Finally, the 1-D spectra for each individual target from all 7 OBs were normalised and combined in a median
stack.

From the reduced spectra, we determine spectroscopic redshifts $z_{\rm spec}$ using the {\sc ez} and
{\sc sgnaps} packages \citep{Garilli_etal:2010,Paioro&Franzetti:2012}.  With {\sc ez}, redshifts are
determined using\break a cross-correlation of spectral templates.  Here we use the default\break templates, but also
include the additional templates of \cite{Bradshaw_etal:2013}, constructed for UDSz.  Optimal solutions were
confirmed via spectral line identification in {\sc sgnaps}.  Overall, the spectra had sufficient S/N to
determine $z_{\rm spec}$ in $\sim70$ per cent of cases.

\vspace{-0.5cm}

\section[]{Results}

\label{Results}

To confirm the nature of our PSB candidates, we determine the presence of strong Balmer absorption lines
(e.g.\ H\,$\delta$ $\lambda4102$\,\AA) combined with a strong Balmer break \citep{Wild_etal:2009}.  An
equivalent width in H\,$\delta > 5$\,\AA\, is generally considered diagnostic of a PSB galaxy
\citep[although often combined with a lack of significant emission lines; e.g.][]{Goto:2007}.
Equivalent width ($W_{\lambda}$) is defined~as
\begin{equation}
{W_{\lambda}} = \int_{\lambda_{1}}^{\lambda_{2}}\!1-F(\lambda)/F_{\rm c}(\lambda)\,{\rm{d}}\lambda,
\end{equation}
where $F(\lambda)$ is the spectral flux and $F_{\rm c}(\lambda)$ is the continuum emission.  We compute the
rest-frame H$\,\delta$ equivalent width ($W_{\rm{H\,\delta}}$) using a non-parametric approach based on that
used by previous works \citep[e.g.][]{Goto_etal:2003}.  First, $z_{\rm{spec}}$ is used to transform the
spectrum into the rest-frame.  Then continuum emission is estimated across H\,$\delta$ using a linear
interpolation between the continuum measured in narrow intervals on either side ($4030$--$4082$ and
$4122$--$4170$\,\AA).  These intervals are chosen for their lack of significant absorption/emission lines
and their continuum is modelled by a linear regression that includes both intervals, weighted by the inverse
square error in the flux.  A $3$-$\sigma$ rejection to deviant points above/below an initial continuum model
is also used to minimise the effect of noise.  Finally, $W_{\rm{H\,\delta}}$ is determined using the
ratio $F(\lambda)/F_{\rm c}(\lambda)$ across an interval that encapsulates H\,$\delta$ ($\lambda_{1} = 4082$
to $\lambda_{2} = 4122$\,\AA).

In this study, the need to identify H\,$\delta$ ($\lambda4102$\,\AA) restricts unambiguous PSB classification
to $z < 1.4$, where H\,$\delta$ is within spectral reach of VIMOS.  This is the case for $24$/$37$ of our PSB
candidates (see Fig.~\ref{Targets}).  In this regime, due to the faintness of our targets the S/N was only
sufficient for a secure $z_{\rm spec}$ and $W_{\rm{H\,\delta}}$ measurement in $15/24$ of cases.  For one of
these cases, an anomaly in the continuum region made the measurement of $W_{\rm{H\,\delta}}$ problematic,
possibly due to a close/projected companion.  This candidate is consequently removed from our sample.  From
the remaining `good' candidates, we find that the majority ($11/14$; $\sim80$ per cent) do exhibit the strong
Balmer absorption ($W_{\rm{H\,\delta}} > 5$\,\AA) and a Balmer break, characteristic of PSB galaxies.  This
result demonstrates the overall effectiveness of the \cite{Wild_etal:2014} PCA technique for selecting
high-redshift post-starbursts at $z < 1.4$.  For the three candidates that were not confirmed, all exhibit
significant but weaker H\,$\delta$ absorption ($W_{\rm{H\,\delta}}\sim3$\,\AA).  Details of our confirmed
PSBs, including their diagnostic $W_{\rm{H\,\delta}}$, are presented in Table~\ref{UDS PSB galaxies}.

\begin{figure}
\includegraphics[width=0.45\textwidth]{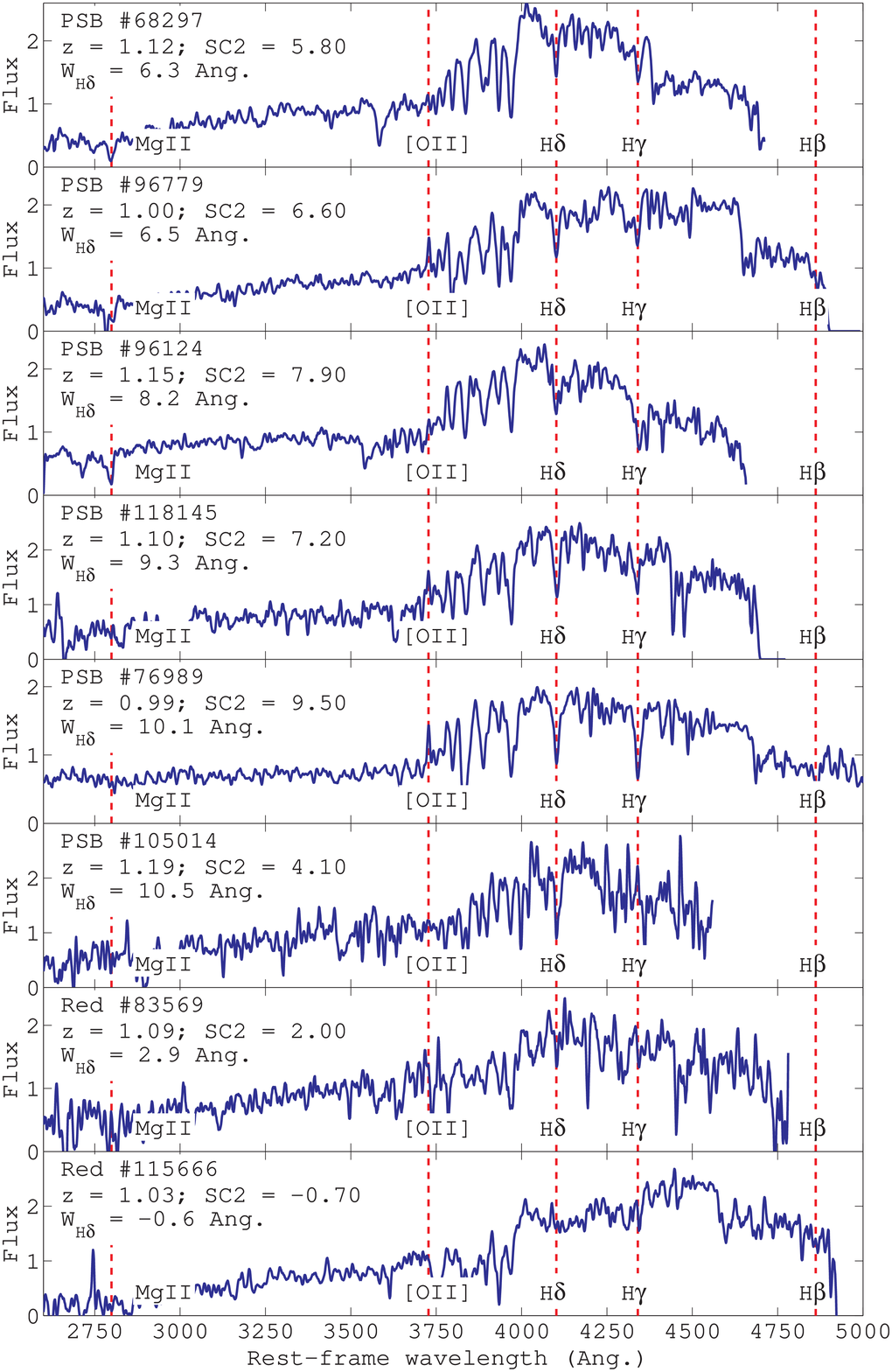}
\centering
\vspace{-0.1cm}
\caption{\label{VIMOS spectra} Example VIMOS spectra for post-starburst (PSB) and red-sequence galaxies in
the UDS.  The PSBs all exhibit the strong Balmer absorption ($W_{\rm{H\,\delta}} > 5$\,\AA) and Balmer break,
characteristic of this population.}
\vspace{-0.0cm}
\end{figure}

We also determine $W_{\rm{H\,\delta}}$ for our red-sequence targets.  Of the cases where a measurement was
possible ($26/34$ galaxies), we confirm that the majority ($18$/$26$; $\sim70$ per cent) exhibit
either very weak or insignificant H\,$\delta$ absorption ($W_{\rm{H\,\delta}} < 5$\,\AA; see
Fig.~\ref{VIMOS spectra}, for examples).  For the remaining cases ($8$ galaxies; $\sim30$ per cent),
post-starburst signatures were actually observed ($W_{\rm{H\,\delta}} > 5$\,\AA; see
Table~\ref{UDS PSB galaxies}).  Some overlap between the populations is not surprising, however, since
photometric methods for identifying PSBs will be less sensitive to post-starburst signatures than using
spectroscopy.

For further verification, we also determine $W_{\rm{H\,\delta}}$ for additional PSB/red-sequence
candidates using the lower-resolution spectra available from UDSz \citep[the spectroscopic component of the
UDS; see][]{Bradshaw_etal:2013,Mclure_etal:2013}.  \cite{Wild_etal:2014} used UDSz for the initial
verification of the SC method, but only used spectra from a limited redshift range ($0.9 < z < 1.2$; $5$
PSBs).  Here, however, we re-analyse the UDSz data using spectra from a wider redshift range
($0.5 < z < 1.4$).  Using only good-quality spectra (S/N $> 7$), we find that the majority ($9$/$11$;
$\sim80$ per cent) of PSB candidates exhibit strong Balmer absorption ($W_{\rm{H\,\delta}} > 5$\,\AA),
whereas the majority ($64$/$70$; $\sim90$ per cent) of red-sequence galaxies do not.  These UDSz results are
in strong agreement with those from our new higher-resolution VIMOS spectroscopy.

\begin{figure}
\includegraphics[height=0.45\textwidth]{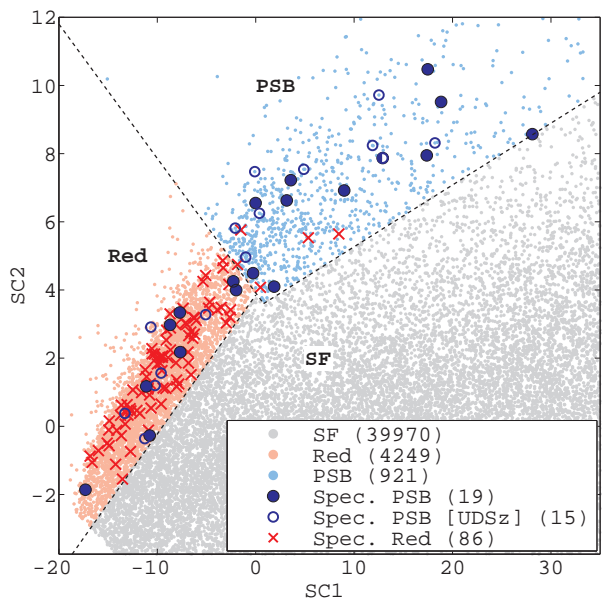}
\centering
\vspace{-0.4cm}
\caption{\label{Supercolour diagram} A super-colour (SC) diagram for galaxies in the UDS field
\mbox{($0.5<z< 2.0$)}, showing our spectroscopically confirmed post-starbursts (PSB).  The
photometrically-selected populations (star-forming, red-sequence and PSB) are indicated by small grey, red
and blue points, respectively.  We also highlight: i)~confirmed PSBs from the latest \mbox{VIMOS}
spectroscopy ($W_{\rm{H\,\delta}} > 5$\,\AA; filled blue circles); ii)~confirmed PSBs from UDSz (open blue
circles); and iii) confirmed red-sequence galaxies ($W_{\rm{H\,\delta}} < 5$\,\AA; red crosses).  Sample
sizes are shown in the legend.}
\end{figure}

In Fig.~\ref{Supercolour diagram}, we show the SC1--SC2 diagram used to identify PSB candidates in the UDS,
and highlight the location of our spectroscopically confirmed candidates (based on
$W_{\rm{H\,\delta}}$ only).  Overall, confirmed PSBs span the full range of SC space used for their
photometric selection, with the low contamination level in this region indicating the general effectiveness
of the demarcations between PSBs, red-sequence galaxies (Red), and star-forming galaxies (SF).
Combining results from our new spectroscopy with UDSz, we confirm that $19/24$ ($\sim80$ per cent) of our
photometrically-selected PSB candidates exhibit PSB spectral signatures (note: there is minimal overlap
between these spectroscopic samples; 1~PSB and 1~Red).  We also determine that the overall completeness in
photometric PSB selection is $\sim60$ per cent, with some spectroscopically confirmed PSBs appearing in the
photometric red-sequence population.  However, as previously noted, this level of completeness is not
unexpected, since the photometric PCA method will naturally be less sensitive to PSB signatures than using
spectroscopy.

\begin{table}
\begin{minipage}{85mm}
\centering
\caption{\label{UDS PSB galaxies} Post-starbursts classified from the new higher-resolution VIMOS
spectroscopy.  Post-starbursts identified from both the SC-selected PSB and red-sequence samples are shown,
corresponding to the filled blue circles in Fig.~\ref{Supercolour diagram}. Errors in
$W_{\rm{H\,\delta}}$ ($1\sigma$) are determined from the $W_{\rm{H\,\delta}}$ variance between simulated
stacks of the $7$ OB spectra generated via a bootstrap method.}
\begin{tabular}{lccccc}
\hline
{DR8 ID}	&{RA}	&{Dec}	&{$z_{\rm spec}$}	&{$K_{\rm AB}$}		&{$W_{\rm{H\,\delta}}$(\AA)}		\\
\hline
{SC-PSB}	&{}		&{}			&{}			&{}		&{}			\\
{$68297$}	&{$34.557\,07$}	&{$-5.218\,42$}		&{$1.122$}		&{$20.2$}	&{$6.3\pm0.3$}		\\
{$68884$}	&{$34.361\,23$}	&{$-5.214\,09$}		&{$0.913$}		&{$22.6$}	&{$9.5\pm3.3$}		\\
{$75927$}	&{$34.368\,46$}	&{$-5.181\,22$}		&{$0.651$}		&{$22.6$}	&{$7.2\pm0.9$}		\\
{$76989$}	&{$34.388\,89$}	&{$-5.176\,62$}		&{$0.987$}		&{$21.4$}	&{$10.1\pm0.4$}		\\
{$96124$}	&{$34.308\,28$}	&{$-5.088\,44$}		&{$1.147$}		&{$20.8$}	&{$8.2\pm0.5$}		\\
{$96779$}	&{$34.584\,38$}	&{$-5.085\,98$}		&{$1.003$}		&{$20.4$}	&{$6.5\pm0.4$}		\\
{$97148$}	&{$34.333\,52$}	&{$-5.084\,09$}		&{$1.273$}		&{$20.5$}	&{$15.2\pm0.4$}		\\
{$102822$}	&{$34.305\,97$}	&{$-5.056\,90$}		&{$0.539$}		&{$20.9$}	&{$8.9\pm0.6$}		\\
{$105014$}	&{$34.395\,62$}	&{$-5.046\,24$}		&{$1.193$}		&{$21.2$}	&{$10.5\pm0.9$}		\\
{$118108$}	&{$34.327\,20$}	&{$-4.984\,63$}		&{$1.398$}		&{$20.8$}	&{$9.6\pm1.6$}		\\
{$118145$}	&{$34.583\,69$}	&{$-4.983\,89$}		&{$1.095$}		&{$21.3$}	&{$9.3\pm0.5$}		\\[1ex]
{SC-Red}	&{}		&{}			&{}			&{}		&{}			\\
{$67085$}	&{$34.422\,45$}	&{$-5.222\,67$}		&{$0.643$}		&{$22.3$}	&{$6.6\pm0.9$}		\\
{$80264$}	&{$34.373\,55$}	&{$-5.163\,23$}		&{$1.091$}		&{$20.0$}	&{$6.3\pm0.8$}		\\
{$83492$}	&{$34.546\,93$}	&{$-5.147\,07$}		&{$1.138$}		&{$20.3$}	&{$8.0\pm0.5$}		\\
{$86288$}	&{$34.584\,42$}	&{$-4.133\,21$}		&{$0.571$}		&{$21.4$}	&{$7.5\pm0.7$}		\\
{$98931$}	&{$34.527\,46$}	&{$-4.076\,77$}		&{$1.270$}		&{$20.1$}	&{$9.8\pm0.8$}		\\
{$105911$}	&{$34.357\,17$}	&{$-5.042\,53$}		&{$1.148$}		&{$20.1$}	&{$5.2\pm0.7$}		\\
{$118886$}	&{$34.341\,84$}	&{$-4.981\,42$}		&{$1.097$}		&{$20.1$}	&{$7.4\pm0.5$}		\\
{$119803$}	&{$34.358\,94$}	&{$-4.976\,03$}		&{$1.024$}		&{$21.8$}	&{$8.1\pm1.7$}		\\
\hline
\end{tabular}
\end{minipage}
\end{table}

With respect to these results, an important consideration is that we do not impose a lack of
[O\,{\sc ii}] emission in our PSB criteria.  [O\,{\sc ii}] emission is not considered typical for PSBs;
however, we do not use this as a classification criterion to avoid a potential bias against PSBs hosting AGN
\citep{Yan_etal:2006}.  In addition, some [O\,{\sc ii}] emission may actually be expected in PSBs where star
formation was not quenched on a rapid timescale.  For our confirmed PSBs, a small number do
exhibit significant [O\,{\sc ii}] emission ($W_{\rm{[O\,{\textsc{ii}}]}} < -10$\,\AA; $3$ galaxies --
$\#68884$, $\#75927$, and one candidate from UDSz).  However, if we exclude these cases, our confirmation
rate for photometrically-selected PSBs only drops to $16/24$ ($\sim65$ per cent).  Furthermore, even using a
much stricter criterion for [O\,{\sc ii}] emission ($W_{\rm{[O\,{\textsc{ii}}]}} < -5$\,\AA;
e.g.\ \citealt{Tran_etal:2003}), we still find that $14/24$ ($\sim60$ per cent) can be confirmed as genuine
PSBs.  From these results, we conclude that the photometric selection of PSB galaxies is highly effective
with, depending on the precise classification criteria used, between $60$--$80$ per cent of the candidates
showing spectral signatures characteristic of this population.

Taken together, our results show that the PCA selection of PSBs is highly successful for
candidates within the magnitude range of our spectroscopic samples ($K_{\rm AB}<23$).  However, we note that
the PCA selection currently yields PSB candidates down to $K_{\rm AB} = 24$ (see Fig.~\ref{Targets}).  At
present, we do not have spectra for these fainter candidates to confirm their nature.  However, based on
their super-colour distributions (i.e.\ SC1, SC2), there is no indication that these fainter candidates are
intrinsically different to those with $K_{\rm AB} < 23$.  Consequently, provided sufficient S/N in the
spectra, we anticipate a similar success rate among this fainter population.

\vspace{-0.3cm}

\section[]{Conclusions}

\label{Conclusions}

We present spectroscopic verification of the selection of high-redshift ($z>0.5$) PSB galaxies from a new
photometric technique, based on the PCA analysis of galaxy SEDs.  Using deep-optical spectroscopy, we
demonstrate the overall effectiveness of this method, with the majority  of PSB candidates targeted
($\sim80$ per cent) exhibiting the strong Balmer absorption ($W_{\rm{H\,\delta}}>5$\,\AA) and Balmer break,
characteristic of this population.  Furthermore, using stricter criteria, excluding those with
significant [O\,\textsc{ii}] emission ($W_{\rm[O\,{\textsc{ii}}]} < -5$\,\AA), we still find a significant
fraction of PSB candidates ($\sim60$ per cent) can be classified as genuine PSBs.  However, we caution that
using the latter criteria may bias against PSBs hosting AGN, or those with residual star formation.  We
conclude that the PCA technique and its application to the UDS field are highly-effective in selecting large
samples of PSBs at high redshift.

The verification presented of the photometric selection of high-redshift post-starbursts will enable a wide
range of science in both the UDS and other deep surveys from large photometric samples.  Such studies will be
the subject of forthcoming publications, including explorations of their large-scale clustering (Wilkinson
et~al., in preparation), environments (Socolovsky et al., in preparation), mass-functions (Wild et al., in
preparation) and morphology (Almaini et al. submitted; Maltby et al., in preparation).

From our new higher-resolution spectroscopy, we also provide a secure sample of $19$
high-redshift PSB spectra for future study.  In addition, our re-analysis of the lower-resolution UDSz
spectra has yielded an additional $14$ PSB spectra, increasing the number of post-starbursts
spectroscopically classified in the UDS field from $5$~to~$33$ galaxies.  A full analysis of the spectral
features for these post-starbursts is beyond the scope of this study, and will be the subject of future
publications.  However, we do make one initial observation.  Although we do observe some PSBs
with significant [O\,{\sc ii}] emission, we observe no significant signs of AGN activity in the spectra of
our post-starburst galaxies (i.e.\ no high-ionisation emission lines, e.g.\ [Ne\,V]).  For some quenching
models, the presence of an AGN is required to linger past the termination phase to sweep up any remaining
gas \citep[e.g.][]{Hopkins:2012}.  At present we find no evidence for lingering AGN activity in our data,
but further work will be necessary to determine the prevalence of obscured and/or low-luminosity
AGN in this important transition population.

\vspace{-0.3cm}

\section[]{Acknowledgements}

This work is based on observations from ESO telescopes at the Paranal Observatory (programmes 094.A-0410 and
180.A-0776).  RJM acknowledges the support of the European Research Council via the award of a Consolidator
Grant (PI McLure).  We thank the staff at UKIRT for their efforts in ensuring the success of the UDS project.

\vspace{-0.3cm}


\bibliographystyle{mnras} \bibliography{DTM_bibtex}
\bsp

\label{lastpage}

\end{document}